\documentclass[final,5p,times,twocolumn,authoryear]{elsarticle}
\usepackage[T1]{fontenc}
\usepackage{graphicx}
\usepackage{amsmath}
\usepackage{nccmath}
\usepackage{amssymb}
\usepackage{xcolor}
\usepackage{url}
\usepackage{CJKutf8}
\usepackage{chemformula}
\usepackage{float}
\usepackage{diagbox}
\usepackage{threeparttable}
\usepackage{subfigure}
\usepackage{hyperref}
\usepackage{geometry}
\hypersetup{colorlinks=true, urlcolor=blue}

\journal{Journal of High Energy Astrophysics}

\begin{document}
\begin{CJK*}{UTF8}{gbsn}
\begin{frontmatter}

\title{Einstein Probe Discovery of an X-ray Flare from K-type Star PM J23221-0301}

\author[sunyatsen,csst]{Guoying Zhao (赵国英)}

\affiliation[sunyatsen]{organization={School of Physics and Astronomy, Sun Yat-Sen University},
            city={Zhuhai}, postcode={519000}, state={Guangdong}, country={China}}
\affiliation[csst]{organization={CSST Science Center for the Guangdong-Hongkong-Macau Greater Bay Area, Sun Yat-Sen University, Zhuhai, 519082, China}}

\author[Berkeley]{Weikang Zheng}
\affiliation[Berkeley]{organization={Department of Astronomy, University of California, Berkeley, CA 94720-3411}, country={USA}}

\author[sunyatsen,csst]{Rong-Feng Shen (申荣锋)}
\cortext[]{E-mail: zhaogy28@mail2.sysu.edu.cn; shenrf3@mail.sysu.edu.cn}

\author[ihep]{Qingcang Shui}
\affiliation[ihep]{organization={Key Laboratory of Particle Astrophysics, Institute of High Energy Physics, Chinese Academy of Sciences}, postcode={100049}, state={Beijing}, country={China}}

\author[naoc]{Dongyue Li}
\affiliation[naoc]{organization={National Astronomical Observatories, Chinese Academy of Sciences},
            city={20A Datun Road, Chaoyang District}, postcode={100101}, state={Beijing}, country={China}}

\author[huake]{Chang Zhou}
\affiliation[huake]{organization={Department of Astronomy, School of Physics, Huazhong University of Science and Technology}, city={Wuhan}, postcode={430074}, state={Hubei}, country={China}}

\author[PMO]{Tianci Zheng}
\affiliation[PMO]{organization={Purple Mountain Observatory, Chinese Academy of Sciences},
            city={Nanjing}, postcode={210023}, state={Jiangsu}, country={China}}
\author[naoc]{Weimin Yuan}  
\author[naoc]{HeYang Liu}

\author[XMU]{Chong Ge}
\affiliation[XMU]{organization={Department of Astronomy, Xiamen University},
            city={Xiamen}, postcode={361005}, state={Fujian}, country={China}}
\author[XMU]{Junfeng Wang}

\author[Berkeley]{Alexei V. Filippenko}
\author[Berkeley]{Thomas G. Brink}

\author[Cal]{Jordan Forman}
\author[Cal]{Mayra Gutierrez}
\author[Cal]{Isabelle Jones}
\author[Cal]{Ravjit Kaur}
\author[Cal]{Naunet Leonhardes-Barboza}
\author[Cal]{Petra Mengistu}
\author[Cal]{Avi Patel}
\author[Cal]{Andrew Skemer}
\author[Cal]{Anavi Uppal}
\author[Cal]{Nicole Wolff}
\author[Cal]{Michele N. Woodland}
\affiliation[Cal]{organization={Department of Astronomy \& Astrophysics, University of California, Santa Cruz, CA 95064-1077}, country={USA}}

\begin{abstract}
Stellar flares are an intense stellar activity that can significantly impact the atmospheric composition of the surrounding planets and even the possible existence of life. During such events, the radiative energy of the star is primarily concentrated in the optical and X-ray bands, with the X-ray flux potentially increasing by tens or even hundreds of times. Einstein Probe (EP) detected a new X-ray transient EP J2322.1-0301 on 27 September 2024. Its spatial localization shows a high positional coincidence with the nearby high proper motion K-type star PM J23221-0301. Follow-up X-ray observations confirmed the flux enhancement of the source, while optical spectroscopic monitoring revealed time-variable features, particularly the disappearance of the H$\alpha$ emission line. This X-ray flare is consistent with a characteristic fast-rise-exponential-decay (FRED) light curve, with a rise timescale of 1.4 ks, a decay timescale of 5.7 ks, and a total duration of $\sim$ 7.1 ks. The peak luminosity in the 0.5--4.0 keV energy band reached $\sim 1.3 \times 10^{31}$ erg s$^{-1}$, with a total energy release of $\sim 9.1 \times 10^{34}$ erg, consistent with the empirical energy correlations observed in magnetic-reconnection-driven stellar flares, as inferred from the multitemperature plasma structure and H$\alpha$-X-ray energy correlation. This discovery underscores EP’s capability in understanding stellar magnetic activity via observing stellar transients.

\end{abstract}


\end{frontmatter}

\section{Introduction}
\label{introduction}
Stellar flares are sudden electromagnetic outbursts characterized by rapid radiation enhancement spanning X-ray to radio wavelengths,  with typical durations ranging from minutes to hours. Stellar flares are widely observed in late-type stars with spectral types F, G, K, and M. The luminosity of stellar flares is typically in the range $10^{26-33}$ erg s$^{-1}$, with total energy on the order of 10$^{28-37}$ erg, and the temperatures reach 10$^{6-8}$ K~\citep{2010ARA&A..48..241B}. As late-type main-sequence stars, K-type stars exhibit magnetic activity levels significantly higher than G-type stars (e.g., the Sun) but lower than M-type red dwarfs. Recent studies suggest that flare frequency and energy release may profoundly influence the atmospheric escape of their planets~\citep{2019ApJ...871..241D}.

These flares are believed to be caused by the rapid release of magnetic energy during the impulsive reconnection of twisted magnetic fields in the star's outer atmosphere~\citep{2011LRSP....8....6S, 2011AJ....141...50W}. When magnetic field lines reconnect, nonthermal electrons are accelerated toward the stellar surface, heating the material and causing it to expand into the magnetic loop. This expansion drives a shock wave which propagates along the loop, increasing the density and temperature of the confined plasma~\citep{1985ApJ...289..414F,1985ApJ...289..425F,1985ApJ...289..434F}. The plasma then cools through soft X-ray radiation. The density, temperature, and decay time of the flare are directly linked to the geometric properties of the magnetic loop \citep{2004A&A...416..733R}. For the Sun and other dwarf stars, more luminous flares reach higher temperatures and last longer.

The soft X-ray light curve of flares generally consists of an exponential decay. A gradual rise or a decay composed of segments with different $e$-folding times can also occur~\citep{1999ApJ...515..746O, 2004A&A...416..733R}. The rise phase of a flare typically represents the rapid release of magnetic energy through the magnetic reconnection process, while the decay phase corresponds to the gradual cooling of the flare's source region via thermal conduction and radiative losses~\citep{2016PASJ...68...90T}. The flare timescale is often interpreted based on one-dimensional hydrodynamic loop models, in which decay phase reflects the conductive and radiative cooling of confined plasma ~\citep{1983ASSL..102..255H,2003AdSpR..32.1057R,2004A&A...416..733R}. However, solar and stellar flares are inherently multi-dimensional phenomena in which multiple loops brighten and evolve successively~\citep{2011LRSP....8....6S}. In this context, the overall flare duration can be characterized by the reconnection timescale rather than by the cooling of a single loop. Numerical simulations of multi-thread flare models have demonstrated that the total flare duration scales with the reconnection timescale, largely independent of the detailed cooling processes~\citep{2017ApJ...851....4R, 2017ApJ...851...91N}.

The H$\alpha$ emission line is the most frequently observed line among solar flare spectroscopic observations~\citep{1984SoPh...93..105I} and similar H$\alpha$ line asymmetries to those in solar flares have also been reported during stellar flares~\citep{1993A&A...274..245H,1994A&A...285..157G,1994A&A...285..489G}. \cite{2018PASJ...70...62H} demonstrated that flares lead to a significant enhancement of the H$\alpha$ emission-line strength and emphasized that variations in the H$\alpha$ line profile are associated with the dynamics of cool gas in the flare region. The H$\alpha$ emission-line enhancement typically indicates an energy transport from the corona to the chromosphere during the flare~\citep{2025ApJ...980..268M}.

In spite of the substantial progress made in recent years, several fundamental scientific questions in stellar-flare research remain unresolved. First, the physical mechanisms underlying the initiation and energy release of flares, particularly in stars with complex or rapidly evolving magnetic fields, are not yet fully understood. Second, the relationship between stellar flares and associated coronal mass ejections (CMEs) is also poorly constrained, especially for stars beyond the Sun, where direct CME detections remain elusive. Researchers have primarily relied on analyzing plasma motions via Doppler shifts in time-resolved spectroscopy to detect stellar CMEs, yet this method has identified only a few candidates~\citep{2022ApJ...933...92C, 2022ApJ...928..180W,2022A&A...663A.140L,2022NatAs...6..241N}. Moreover, how flare properties such as energy, duration, and recurrence rate vary across different stellar types and evolutionary stages remains an open question. Another key challenge lies in characterizing the flare-induced high-energy radiation environment and assessing its cumulative impact on the habitability and atmospheric retention of exoplanets. Addressing these issues requires a combination of high-cadence, multiwavelength observations and advanced theoretical modeling~\citep{2024Univ...10..313V,2020MNRAS.494.3766O,2025A&A...694A.161S, 2022MNRAS.509.5858H, 2021NatAs...5..298C}. 

Recent studies of stellar flares have utilized the photometric light curve from space surveys, such as the Kepler and K2 missions of the Kepler Space Telescope, and the Transiting Exoplanet Survey Satellite (TESS)~\citep{2014ApJS..211....2H, 2020AJ....159...60G}. The Einstein Probe (EP), launched on 2024 January 9, is dedicated to monitoring the sky in the soft X-ray band~\citep{2022hxga.book...86Y}. Its wide-field X-ray monitoring capabilities have significantly enhanced flare detection efforts, enabling the discovery of an increasing number of stellar flares and thus offer unprecedented opportunities to investigate the energetics and temporal evolution of flares~\citep{2025SCPMA..6839501Y}.

While numerous stellar flares have been reported from previous missions such as Kepler, TESS and XMM-Newton, EP offers a unique capability through its wide-field X-ray monitoring and rapid follow-up focusing mode, enabling the detection and characterization of a stellar flare by EP. This event provides a proof-of-concept demonstration of EP's potential for systematically exploring stellar magnetic activity.

In this paper, we present a comprehensive multiwavelength dataset of the transient EP J2322.1-0301, acquired through a coordinated campaign employing the KAIT photometry and Shane/Kast spectroscopy at Lick Observatory. The paper is structured as follows. We describe the observation and data reduction in Section~\ref{sec:Observation}. In Section~\ref{sec:results}, we analyze the radiation characteristics and temporal evolution of the transient source. Our conclusions are presented in Section~\ref{sec:conclusion and discussion}.

\section{Observations and data reduction}\label{sec:Observation}

Equipped with two scientific instruments, the Wide-field X-ray Telescope (WXT) and the Follow-up X-ray Telescope (FXT), EP offers a large instantaneous filed of view for detecting rapid transients, along with a considerable effective area crucial for follow-up observations and precise localization. EP/FXT operates in the 0.5--10.0 keV X-ray band. It consists of two modules (FXT-A and FXT-B), each containing 54 nested Wolter-I paraboloid-hyperboloid mirror shells. The PN-CCD of FXT offers three readout modes, Full Frame mode (FF), Partial Window mode (PW), and Timing mode (TM).

Subsequently, the WXT photon event data processing and calibration pipeline was applied using dedicated data-reduction software coupled with a calibration database system. The extraction and analysis of time-resolved spectra and light curves were conducted utilizing Fxtsoftv1.05\footnote{\url{http://epfxt.ihep.ac.cn/analysis}} of the EP data analysis software. 

\subsection{Einstein Probe Discovery}
\label{Sec:Einstein Probe Discovery}
\begin{figure}[ht!]
\centering
\includegraphics[width=3.4in]{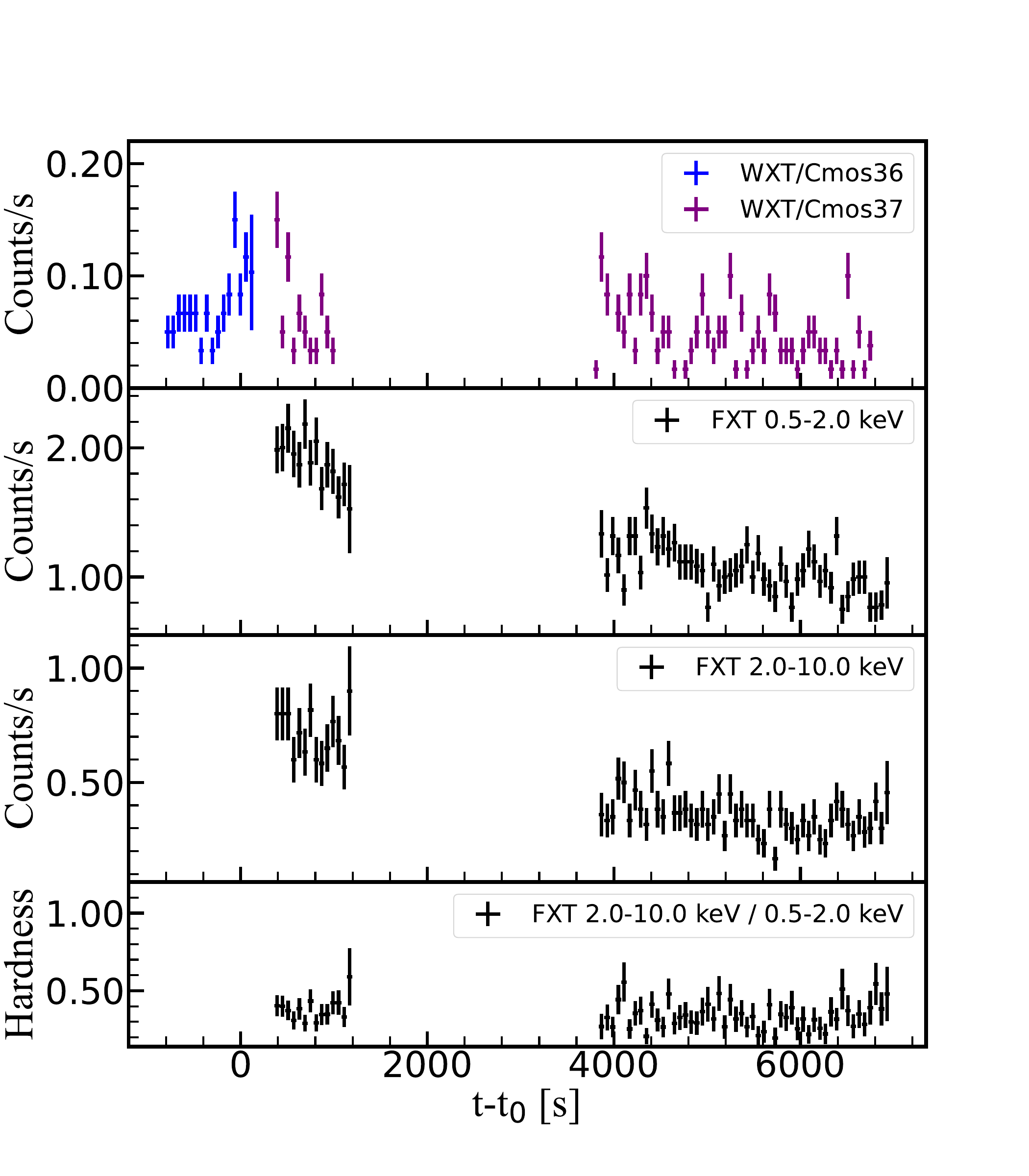}
\caption{The 60~s time-binned light curves of EP J2322.1-0301, as observed by WXT in the 0.5--4.0 keV band and by the FXT in two energy bands, along with the hardness ratio defined as C$_{\mathrm{2.0-10~keV}}$/C$_{\mathrm{0.5-2.0~keV}}$ (exhibiting only slight variations). Here, $t_0$ is the WXT trigger time 2024-09-27T01:03:48 (UTC). WXT/CMOS36 and WXT/CMOS37 represent two adjacent CMOS detection units of EP/WXT, and the target is located in the overlapping region of the fields of view of the two units.}
\label{fig:curve}
\end{figure}

On September 27, 2024 (UTC), EP-WXT detected a new X-ray transient (trigger ID 01709061302, CMOS36)~\citep{2024GCN.37615....1S}, which was designated as EP J2322.1-0301. EP/WXT has a total of 48 detection units (CMOS01-CMOS48), among which CMOS36 and CMOS37 are the two adjacent detection units covering the target sky area in this observation. Since the target is located in the overlapping region of the two units, it is simultaneously detected by CMOS36 and CMOS37. Figure~\ref{fig:curve} presents the temporal evolution of flares observed by WXT in the 0.5-4.0 keV band, as well as the temporal evolution of flares from the FXT in two energy bands (0.5-2.0 keV and 2.0-10.0 keV), along with the evolution of the FXT hardness ratio. For reference, $t_0$ denotes the WXT trigger time of 2024-09-27T01:03:48 (UTC). The light curve of this transient exhibits a typical evolutionary pattern of rapid rise and gradual decay. Notably, the decay phase of the flares is relatively prolonged; meanwhile, the FXT hardness ratio shows only minor variations.

\begin{figure}
\includegraphics[width=3.4in]{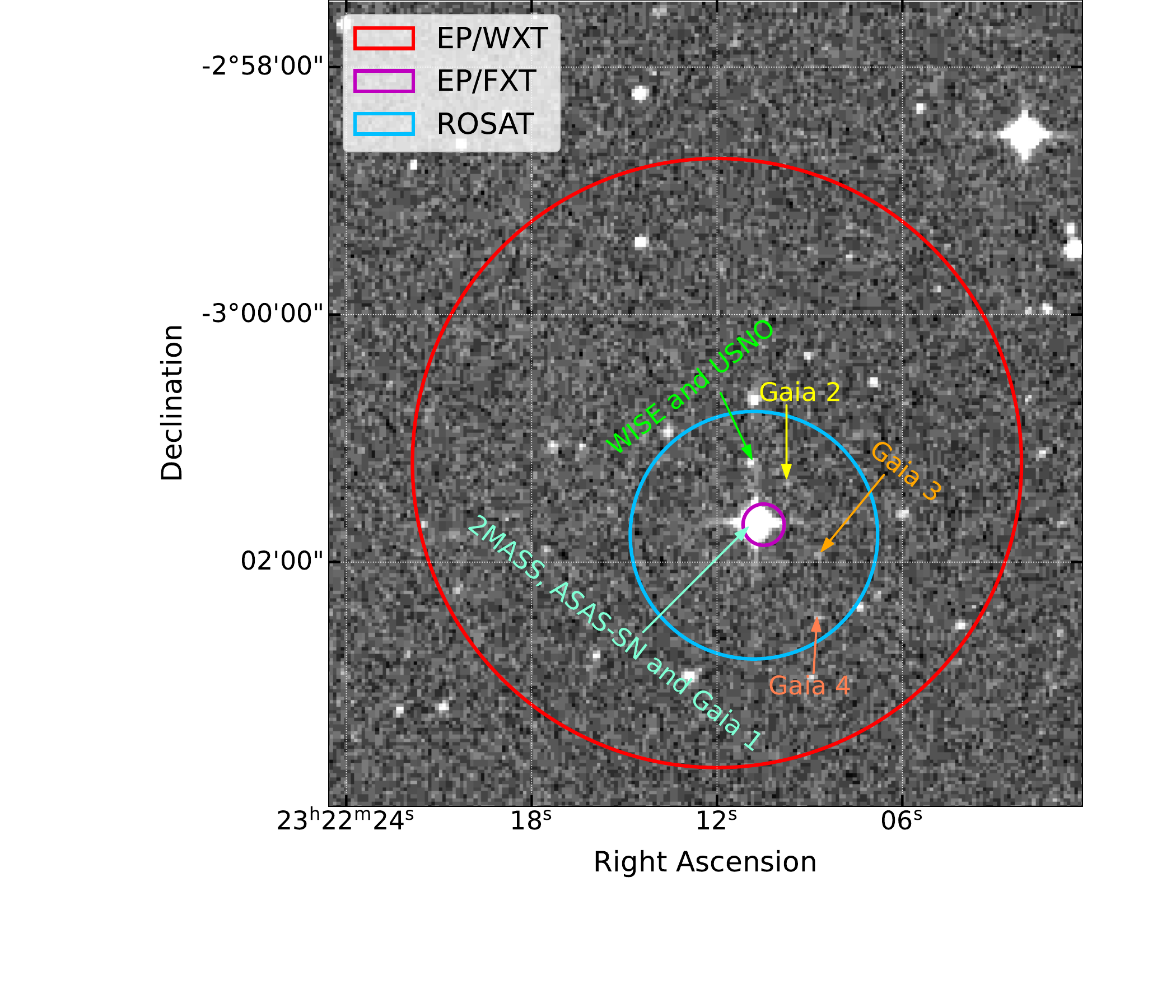}
\caption{The EP/WXT detection localization (red circle, 2.46$^\prime$ error radius), the subsequent EP/FXT refinement (cyan circle, 20$^{\prime\prime}$ radius), and the ROSAT cataloged source localization (blue circle) overlaid on the Digital Sky Survey (DSS) optical image. Potential candidate multiwavelength counterparts within the WXT error circle, including Gaia DR3, 2MASS, WISE, and USNO are also labeled. The high proper motion K-type star PM J23221-0301 is the centrally brighter star indicated by the aquamarine arrow, showing $<6^{\prime\prime}$ offset from the FXT centroid. } 
\label{fig:location}
\end{figure}

The WXT source is positioned at R.A. = 350.551$^\circ$, Dec. $= -3.026^\circ$, with an uncertainty of $2.461'$. FXT initiated follow-up observations 329~s later, pinpointing the source at R.A. = 350.5437$^\circ$, Dec. $= -3.0283^\circ$, with a positional uncertainty of $\sim 10''$, consistent with the WXT transient's location, shown in Figure~\ref{fig:location}. 

Within the WXT error circle, archival data from ROSAT, Gaia DR3, 2MASS, ASAS-SN, WISE, and USNO-B1 were cross-matched. Table~\ref{table:sources} presents the cataloged known sources within this region. Additionally, within the FXT error circle, Gaia, 2MASS and ASAS-SN sources are identified, which are associated with the high proper motion star PM J23221-0301~\citep{2020yCat.1350....0G}. Notably, although other potential counterparts exist within the WXT error circle, all of them are located outside the FXT error circle, exhibiting extremely low association probabilities with the event, compared to PM J23221-0301. This finding strengthens the case for this high proper motion star being the most likely physical host of EP J2322.1-0301.

PM J23221-0301 is located at R.A. = 23:22:10.89, Dec. = -03:01:41.98. Previous ROSAT observations identified it as the X-ray source 1RXS J232210.8-030147, with a quiescent flux of $1.04 \times 10^{-12}$ erg s$^{-1}$ cm$^{-2}$~\citep{1999yCat.9010....0V}, while ASAS-SN optical monitoring revealed sporadic brightening episodes\footnote{\url{https://asas-sn.osu.edu/}}.

Further constraints on the fundamental stellar properties of PM J23221-0301 are derived from its observed atmospheric parameters and archival data. Specifically, SIMBAD\footnote{\url{https://simbad.cds.unistra.fr/simbad/}}  provides a distance measurement of $46.218 \pm 0.0378$ pc~\citep{2020AJ....160...83S}, a stellar rotation period of 1.28 days~\citep{2013AcA....63...53K} and a stellar age of 1.2 Gyr. The star is of K type with effective temperature  $T_\mathrm{eff} = 4055$ K~\citep{2022ApJS..259...35A}. Utilizing the mass-effective temperature scaling relation for K-type main-sequence stars presented in \cite{2013ApJS..208....9P}, we derive a stellar mass of $M_*\approx 0.70~M_\odot$. Combining this with the measured surface gravity (log $g$ = 4.634 [cgs],~\cite{2022ApJS..259...35A}), we calculate the stellar radius to be $R_*\approx0.72~R_\odot$.

\begin{table*}
\caption{Cataloged known sources within the WXT localization error circle.}\label{table:sources}
\centering
\begin{threeparttable}          
    \begin{tabular}{*7{c}}
    \hline 
    \hline
    Instruments &Source Name & R.A.,~Dec.& Distance&Flux&Magnitude&Wave Band\\
    & & & [pc] &[erg cm$^{-2}$ s$^{-1}$]&[mag]&\\
    \hline 
    \hline
    2MASS$^{\mathrm{a}}$& 23221088-0301417 &350.543, -3.028&  &-&8.73&Infrared\\
    ASAS-SN$^{\mathrm{b}}$& J232211-0301.7&350.545,-3.028 &46.2 & -&11.42&Optical\\
    Gaia DR3$^{\mathrm{c}}$& 2637463382668201856 &350.545, -3.028& & -&10.78&Optical\\    
    \hline 
    \hline
    Gaia DR3$^{\mathrm{c}}$& 2637464134287151488 &350.541, -3.022&5058.2 &-&20.10&Optical \\
    \hline 
    \hline
    Gaia DR3$^{\mathrm{c}}$& 2637464099929200256 &350.546, -3.041&364.6&-&20.70&Optical  \\
    \hline 
    \hline
    Gaia DR3$^{\mathrm{c}}$& 2637463344013160832 &350.536, -3.041&- &- &20.6&Optical \\
    \hline 
    \hline
    WISE$^{\mathrm{d}}$& J232210.94-030110.9 &350.546, -3.020&- &-&16.8&Infrared \\
    USNO-B1$^{\mathrm{e}}$& 0869-0701355 &350.546, -3.020&- &- &20.90&Optical \\
    \hline 

    \hline
    ROSAT$^{\mathrm{f}}$& 1RXS J232210.8-030147  &350.545, -3.030& 53  &1.04e-12 &-&X-ray\\

    \hline 

    \hline
    
    \end{tabular} 
            \begin{tablenotes}  
    \footnotesize              
    \item[a] ~\cite{2003yCat.2246....0C}. ${\mathrm{^{b}}}$~\cite{2013yCat.120630053K}. ${\mathrm{^{c}}}$ ~\cite{2022yCat.1355....0G}. ${\mathrm{^{d}}}$~\cite{2012wise.rept....1C}. ${\mathrm{^{e}}}$ ~\cite{2003AJ....125..984M}. ${\mathrm{^{f}}}$ ~\cite{1999yCat.9010....0V}.
    \end{tablenotes}            

\end{threeparttable}  
\end{table*}

\subsection{Lick Observatory KAIT Photometry and Kast Spectroscopy}
\begin{figure}
\centering
\includegraphics[width=3.4in]{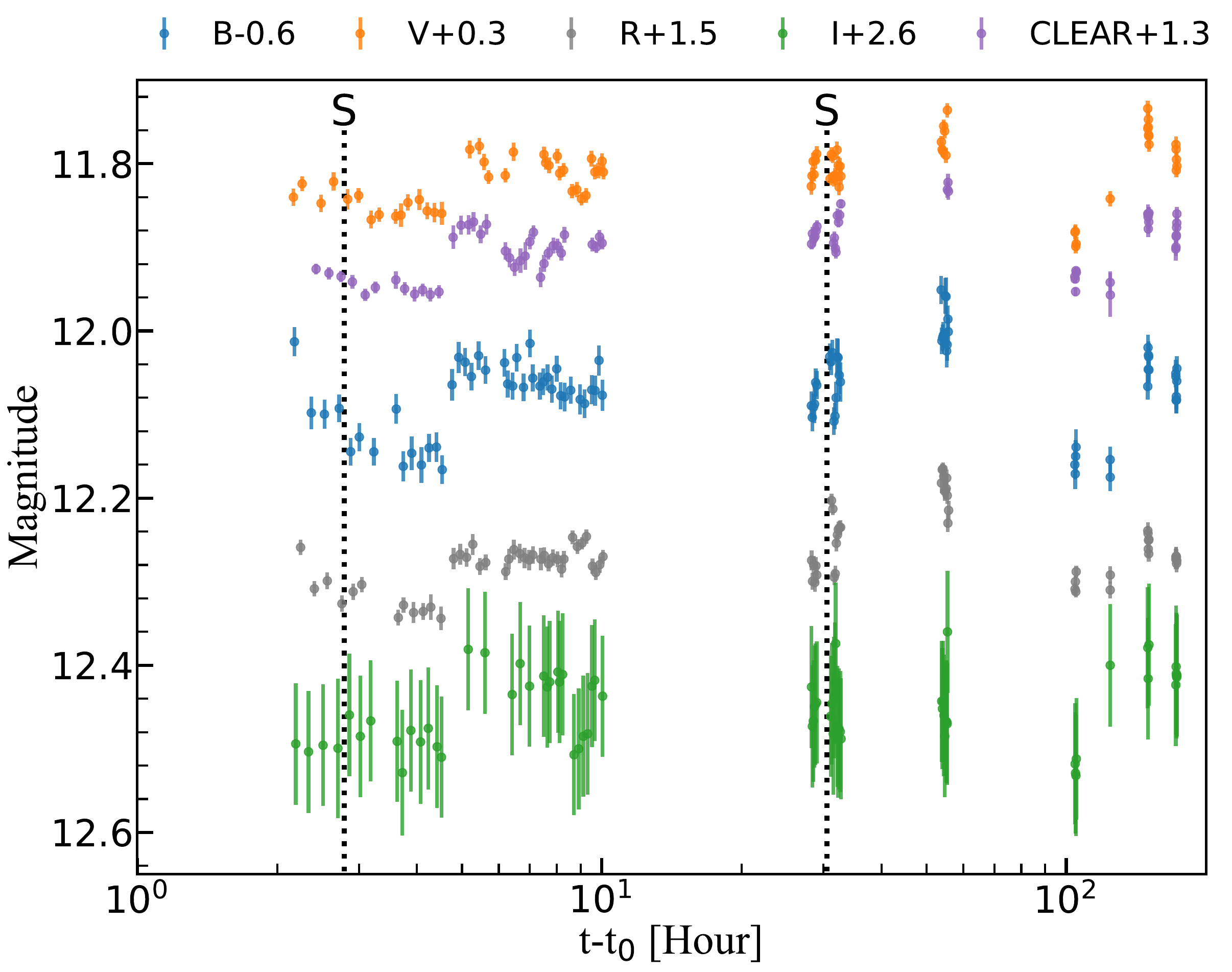}
\caption{Short-term KAIT optical light curves of PM J23221-0301. The ``S'' notations and dash lines mark the commencement of Kast's two spectral observations, corresponding to 2.79 hr and 30.53 hr post-WXT trigger.} 
\label{fig:KAIT_curve}
\end{figure}

Follow-up photometric observations of PM J23221-0301 within the FXT localization region were conducted using the 0.76~m Katzman Automatic Imaging Telescope (KAIT), as part of the Lick Observatory Supernova Search (LOSS;~\citealt{2001ASPC..246..121F}). Observations began $\sim 2$~hr after $t_0$ and continued over subsequent nights to monitor potential post-flare evolution. The KAIT observations utilized standard $B$, $V$, $R$, and $I$ filters, and additional clear-band images~\citep{2011MNRAS.412.1441L}. The resulting optical light curves shown in Figure~\ref{fig:KAIT_curve} 
exhibit no clear evidence of flaring activity, although low-amplitude fluctuations ($\sim$0.2 mag) are present across all photometric bands. This absence of activity may be due to the delayed start of the KAIT observations, which likely missed some prompt activity associated with the initial high-energy transient signal.

\begin{figure}[ht!]
\centering
\includegraphics[width=3.7in]{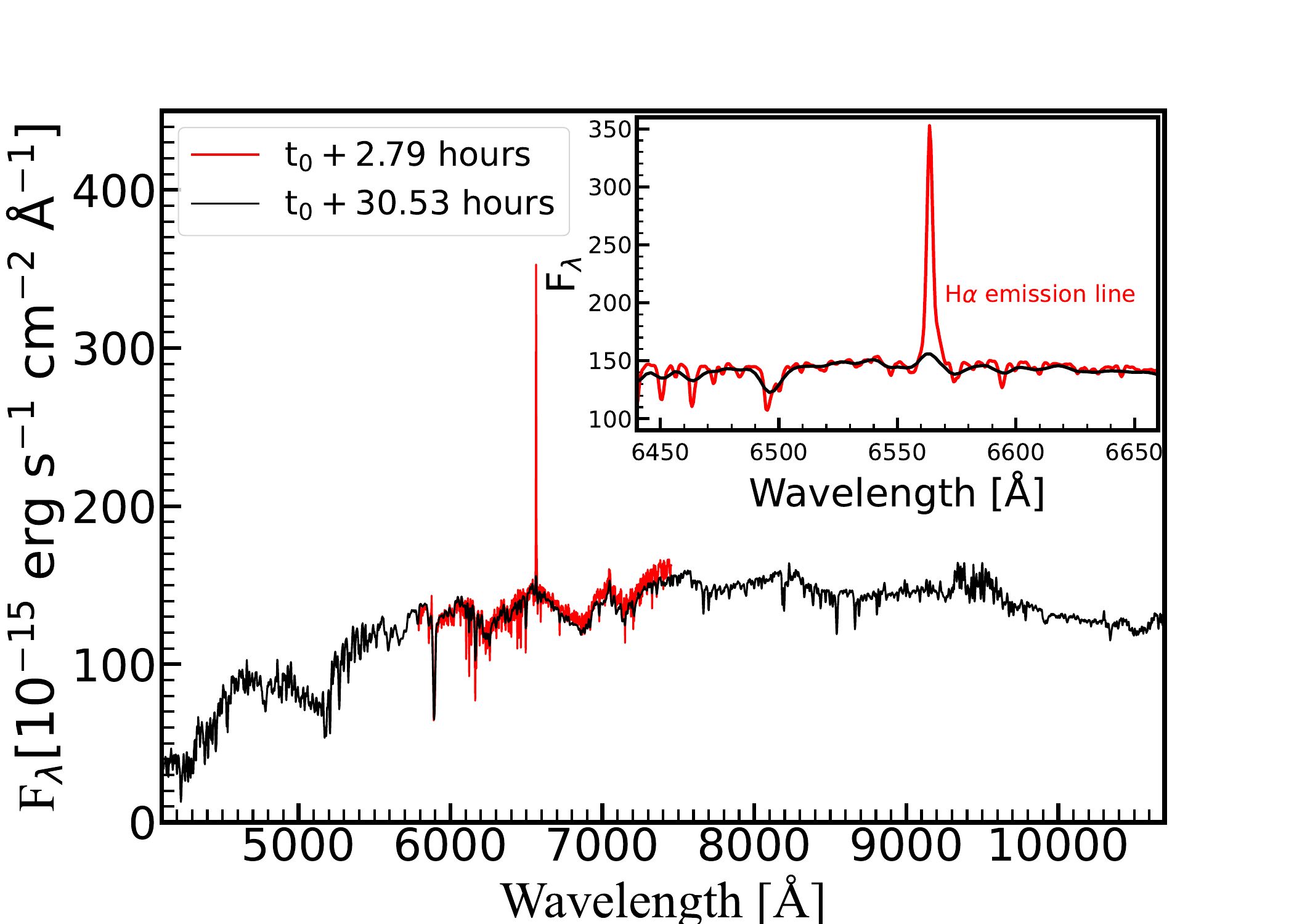}
\caption{Optical spectra of PM J23221-0301, obtained using Kast on two consecutive nights: 2.79 hr (red) and 30.53 hr (black) post-WXT trigger.  Notably, as shown in the zoomed-in panel, the 2.79 hr spectrum features a distinct H$\alpha$ emission line ($\lambda= 6563$ $\AA$); however, this emission line is absent the next night.} 
\label{fig:KAIT_spectrum}
\end{figure}

Optical spectra of PM J23221-0301 were obtained using the Kast spectrograph on the Shane 3~m telescope at Lick Observatory~\citep{1988igbo.conf..157M}. The instrumental resolution is $\sim 18 \, \AA$ at the central wavelength of 6800 $\AA$. The first epoch observation was conducted on 2024-09-27T03:50:24 ($t_0$+2.79 hr), covering a wavelength range of 5800--7400 $\text{\AA}$~\citep{2025GCN.39707....1Z}. As shown in  Figure~\ref{fig:KAIT_spectrum}, the spectrum shows a red continuum with narrow absorption lines consistent with a K-type star spectrum. Notably, a strong H$\alpha$ emission line is present, confirming the active nature of PM J23221-0301. The H$\alpha$ line profile in the first epoch is symmetric within the spectral resolution, showing no evident blue- or red-wing asymmetry.

To investigate the temporal evolution of this activity, a second-epoch spectrum was obtained on the following night, $\sim 1.272$ days after the trigger, covering a broader wavelength range of 3500--10,000 $\text{\AA}$~\cite[see Figure~\ref{fig:KAIT_spectrum}]{2025GCN.39707....1Z}. In this spectrum, the H$\alpha$ emission line had nearly disappeared compared to the previous spectrum, confirming PM J23221-0301 as the optical counterpart of the EP flare event.

\subsection{ASAS-SN}

\begin{figure}[ht!]
\centering
\includegraphics[width=3.4in]{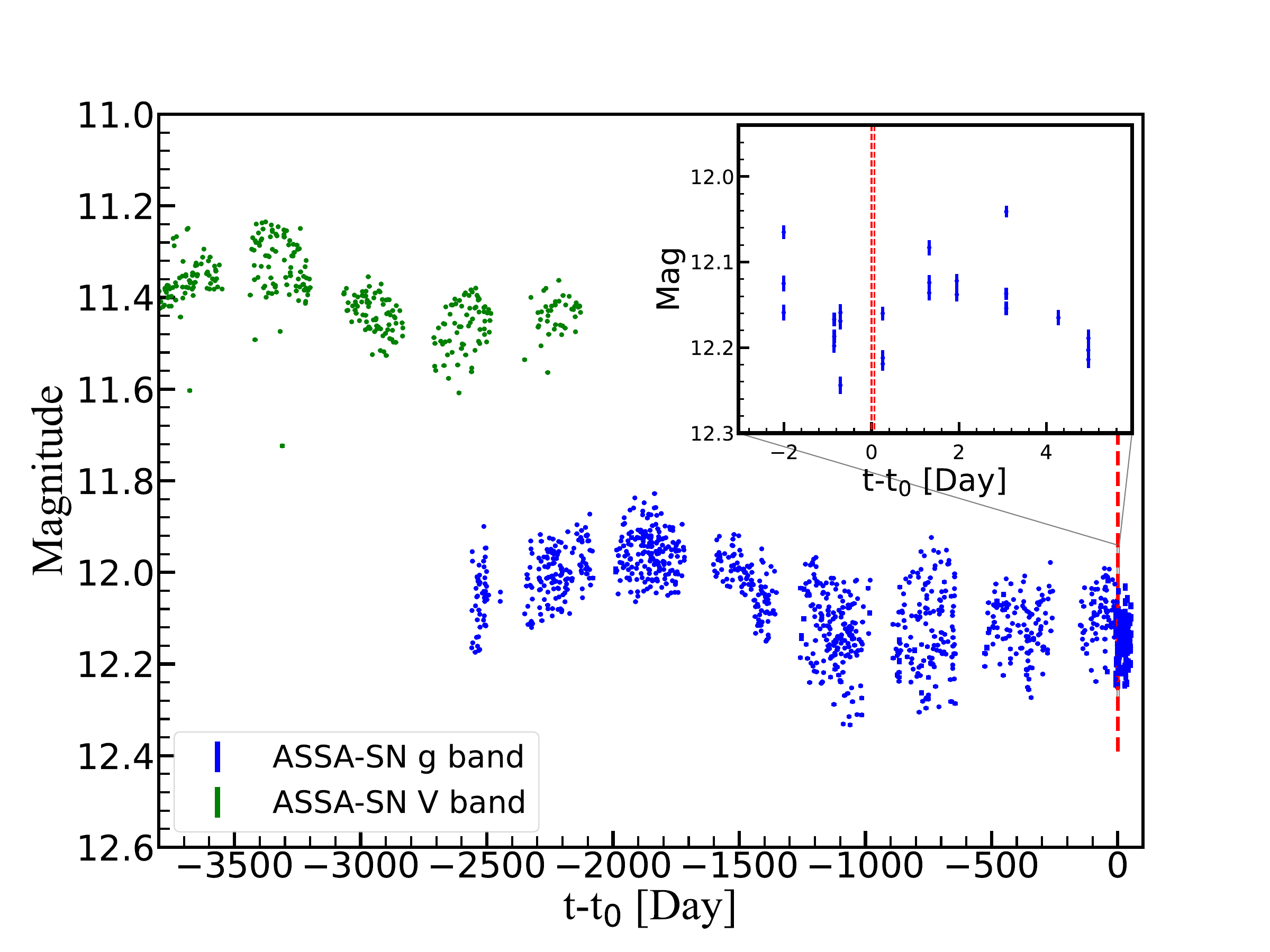}
\caption{Long-term ASAS-SN optical light curves of PM J23221-0301, with data obtained from \protect\url{https://asas-sn.osu.edu/}. The red dashed vertical lines demarcate the EP observation window.} 
\label{fig:ASSA-SN_curve_with_inset}
\end{figure}

PM J23221-0301 is also regularly monitored in the optical by the ASAS-SN project~\citep{2014ApJ...788...48S}. While no significant outburst was detected during the FXT observation window, a relatively small variation in the optical light curve was identified between $t_0-1$ and $t_0+5$ days, shown in Figure ~\ref{fig:ASSA-SN_curve_with_inset}. The absence of pronounced optical flaring signatures may be attributed to the temporal resolution limitations of the ASAS-SN monitoring, whose sampling cadence is $\sim 1$ day. Given the transient nature of the X-ray emission (duration $\sim 7.1$ ks), the rapid flux variations could have been averaged out below the optical detection threshold. This likely prevented the capture of any optical counterpart synchronous with the X-ray flare.

\section{Spectral and Temporal Analysis}\label{sec:results}

\subsection{X-ray Spectral Fitting}\label{sec:X-ray_spectral_fitting}

Stellar flares are typically emitted from a hot plasma triggered by magnetic reconnection, which exhibit radiation characteristics dominated by thermal continuum emission superimposed with metal-line emission. In X-ray astronomy, the emission from hot plasmas is typically described by the Astrophysical Plasma Emission Code \citep[APEC;][]{2001ApJ...556L..91S} model with two free parameters: the normalization of surface brightness and the temperature. The APEC model accurately simulates the spectral features of thermal plasma by computing ionization equilibrium, atomic transitions, and line-emission processes, making it particularly suited for spectral fitting of hot coronal regions or magnetic loop structures during flares~\citep{2001ApJ...556L..91S,2005HiA....13..651B}. 

For a hot plasma at an angular distance of $D_A$ to the observer, the observed spectral flux $F(E)$ $\mathrm{[erg~cm^{-2}~s^{-1}~keV^{-1}]}$ is obtained by integrating over the plasma volume,
\begin{equation}
    {F(E)}={\frac{1}{4\pi[D_A(1+z)]^2}}\int_V \Lambda(E,T)~dV,
\end{equation}
where  $z$ is the redshift, and $\Lambda(E,T) \propto n_e n_H T^{-1/2}e^{-E/(kT)}$ is the photon emissivity $\mathrm{[erg~cm^{-3}~s^{-1}~keV^{-1}]}$ for bremsstrahlung (thermal continuum radiation), which depends on the plasma temperature $T$ and photon energy $E$ ($n_e$ and $n_H$ are the electron and hydrogen number densities [cm$^{-3}$], respectively). For practical use, APEC uses a normalization factor,
\begin{equation}
    {\rm norm}=\frac{10^{-14}}{4\pi{[D_A(1+z)]^2}}\int_V {n_e n_H dV}.
\end{equation}

\begin{figure}[ht!]
    \centering
    \subfigure[]{
        \includegraphics[width=3.0in]{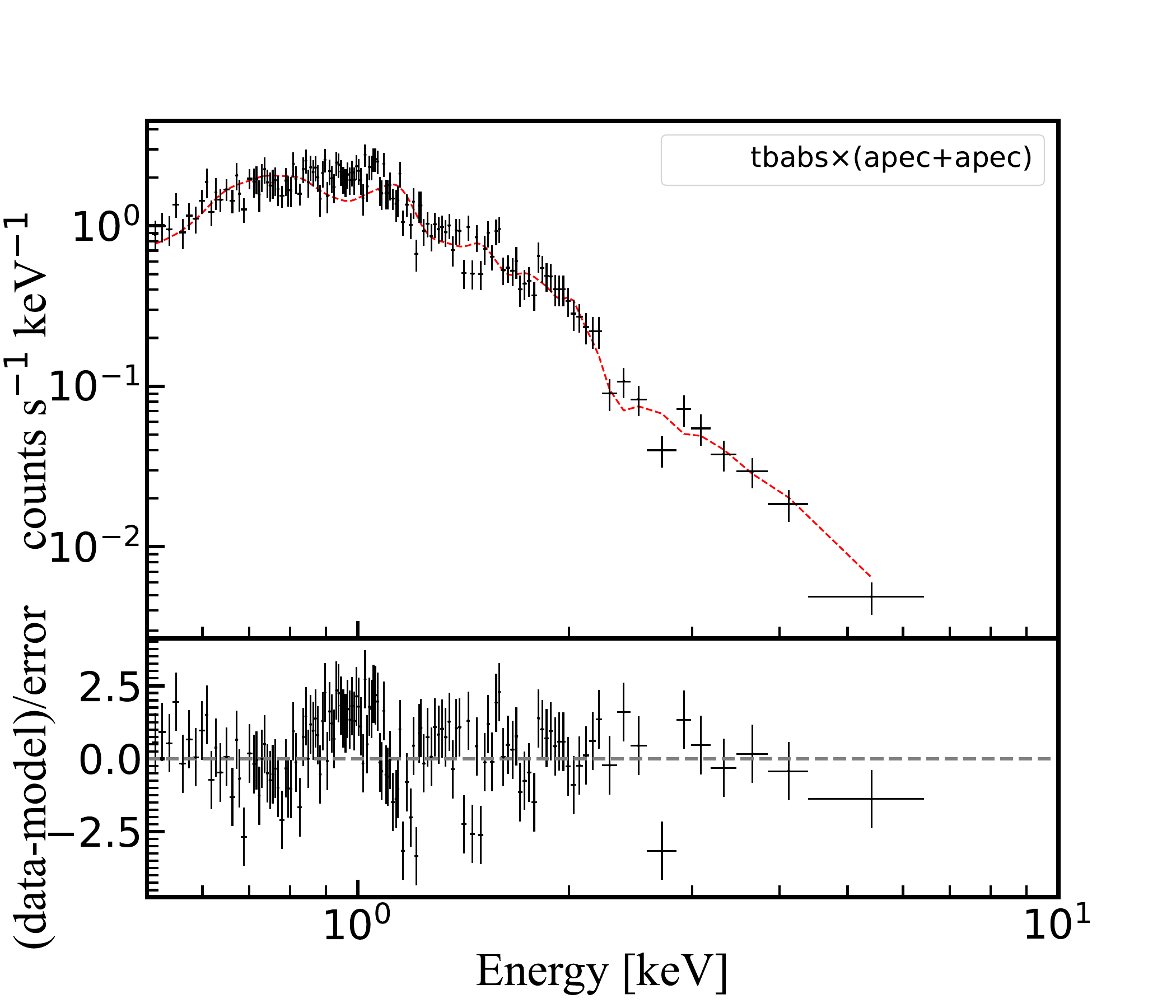}
    \label{fig:2apec_count}}
    \subfigure[]{
	\includegraphics[width=3.0in]{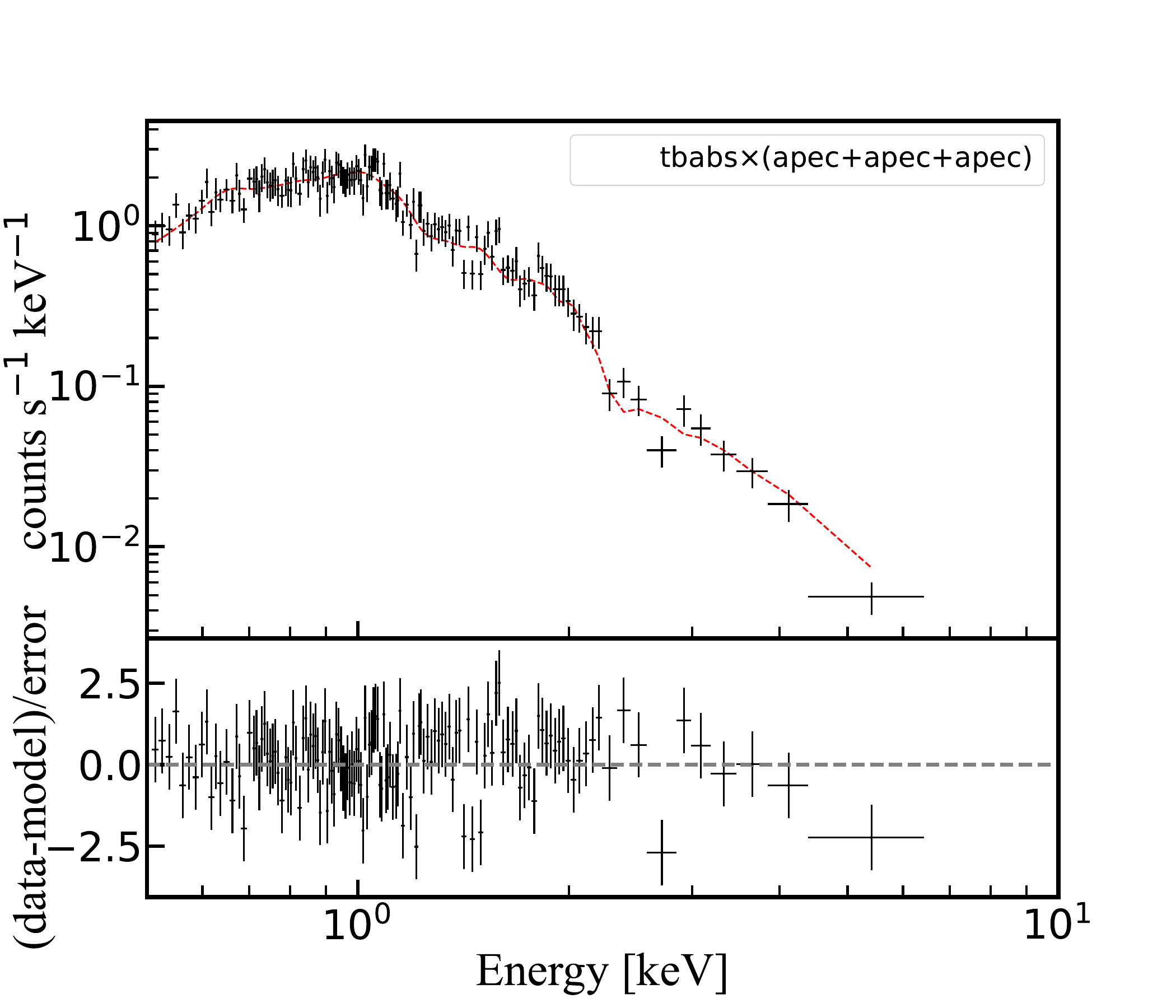}
    \label{fig:3apec_count}}    
    \caption{Spectral fit of the flare in the 0.5--10.0 keV band from t$_0$+329s to t$_0$+6929 s. The top section of each panel is the observed energy spectrum and the model (red dotted line), while the bottom section shows the residuals. Panel (a) is for a two-APEC model, which exhibits significant residuals at $E \approx 1$ keV. Panel (b), utilizing a three-APEC model with an additional thermal component, effectively improves the fitting quality in the 1 keV region, with residuals notably reduced. This indicates that a multitemperature model can more accurately characterize this flare.} 
    \label{fig:spec_res_fit}
\end{figure}

We first performed a spectral analysis of the time-integrated FXT data over the entire FXT observation period of $t_0+$ (329 s, 6929 s) via  Xspec\footnote{\url{https://heasarc.gsfc.nasa.gov/xanadu/xspec/manual/manual.html}}. For the spectral model, we initially tried a two-temperature collisional ionization equilibrium model ($\text{tbabs} \times (\text{apec} + \text{apec})$). The first component is responsible for the Galactic absorption using the Tuebingen-Boulder interstellar medium absorption model~\citep{Wilms_2000}. For the Galactic hydrogen column density, we adopted $N_H = 1.01 \times 10^{16}$~cm$^{-2}$, as calculated by  Xspec. This model yielded a reduced chi-squared of $\chi^{2}/{\rm dof} =222.59/133$, with significant residuals observed at low energies (see Figure ~\ref{fig:spec_res_fit}). Table \ref{tab:apec_para} summarizes the results of the fitting and the derived parameters, offering a clearer evaluation of the quality of fitting.

To address this low-energy excess residual in fitting, we introduce an additional thermal component, constructing a three-temperature model ($\text{tbabs} \times (\text{apec} + \text{apec}+ \text{apec})$). For the Galactic hydrogen column density, we adopt the value derived from the two-component APEC model fit. The implementation of the three-APEC model 
significantly improved the fitting quality, yielding $\chi^{2}/{\rm dof}=143.29/131$, as shown in Table \ref{tab:apec_para}, which demonstrates a statistically significant improvement compared to the two-APEC model. This implies the necessity of multitemperature plasma components to characterize this flare event. 

\begin{table*}
\caption{\label{tab:apec_para}Best-fitting parameters for the FXT spectra.}
\centering
\renewcommand{\arraystretch}{1.3}
\begin{tabular}{cccccc}
\hline 
\hline

&${kT{_1}}$&${kT{_2}}$ &${kT{_3}}$&$F_{0.5-10~\rm keV}$  &$\chi^{2}$/dof\\
&[$\mathrm{keV}$]&[$\mathrm{keV}$]&[$\mathrm{keV}$]&[$10^{-11}$~erg~s$^{-1}$~cm$^{-2}$]&\\

\hline
 
$\mathrm{tbabs\times (apec+apec)}$ & $3.40_{-0.32}^{+0.40}$ &$0.32_{-0.04}^{+0.03}$&-&$1.22_{-0.20}^{+0.16}$ &222.59/133 \\

$\mathrm{tbabs\times(apec+apec+apec)}$& $4.71_{-0.73}^{+1.18}$&$1.12_{-0.11}^{+0.12}$& $0.27_{-0.03}^{+0.03}$&  $1.36_{-0.01}^{+0.06}$&143.29/131\\ 

\hline
\hline 
\end{tabular}
\end{table*}

Next, we use the Xselect tool to perform  time binning of the WXT and FXT data, respectively, and then extract the spectral data in the 0.5--4.0 keV energy range for each time slice. In the rise phase of the flare, varying time bins were applied to Cmos36 (413 s, 430 s and 117 s), Cmos37 and FXT. Given that the Cmos37 and the FXT observing time periods are mostly overlapping, we jointly fitted the data from the two instruments to obtain more accurate fluxes. During the decay phase, Cmos37 and FXT adopted a $\sim$500 s time bin to continuously acquire 7 data points. Since these sampling intervals are significantly smaller than the flare's rise and decay timescales, the flux evolution process was effectively captured. Background subtraction was executed with source-centered annular regions. 

We then independently fitted each time slice with the three-temperature APEC model via Xspec software to derive the absorption-corrected X-ray fluxes, with the absorption set to the Galactic value ($N_H=1.01\times10^{16}~\mathrm{cm}^{-2}$ ). The  results are presented in Table~\ref{table:time_resoul}. The temperature of the hot component varies significantly over time, whereas the temperature of the cool component remains relatively stable throughout the observation period. 

\begin{table*}
\caption{Time-resolved spectral fitting results.}\label{table:time_resoul}
\centering
\begin{threeparttable} 
\renewcommand{\arraystretch}{1.3} 
    \begin{tabular}{*7{c}}
    \hline 
    \hline
    Instruments &Time Interval & $kT_1$& $kT_2$& $kT_3$ &Flux&Cstat/dof\\
    &[s]& [keV]& [keV] & [keV]  &[$10^{-11}$ erg cm$^{-2}$ s$^{-1}$]&\\
    \hline 
    \hline
    & (-843, -430) & 6.42$_{-1.13}^{+1.40}$ &1.25$_{-0.19}^{+0.20}$&1.07$_{-0.10}^{+0.10}$&1.7$\pm 0.1$&10/9\\
    WXT& (-430, 0) & 6.81$_{-1.90}^{+2.50}$ &1.22$_{-0.25}^{+0.15}$&0.87$_{-0.11}^{+0.21}$&2.5$\pm 0.2$&8.8/6\\   
    (cmos36)& (0, 117) & 6.96$_{-1.50}^{+2.21}$&1.24$_{-0.10}^{+0.10}$&1.01$_{-0.10}^{+0.10}$&4.9$\pm 0.5$&6.7/4\\
    \hline 
    \hline
    & 329-659 & 6.61$_{-0.92}^{+1.24}$ &1.24$_{-0.07}^{+0.11}$&0.27$_{-0.02}^{+0.01}$&5.0$\pm 0.4$&235.5/176\\
    & 659-989 & 6.85$_{-1.01}^{+1.01}$ &1.02$_{-0.05}^{+0.05}$&0.27$_{-0.01}^{+0.01}$&4.8$\pm 0.2$&224/191\\   
    & 3869-4379 & 3.77$_{-0.49}^{+0.49}$&1.48$_{-0.15}^{+0.15}$&0.95$_{-0.03}^{+0.07}$&2.7$\pm 0.2$&250.9/195\\
    WXT (cmos37)+FXT& 4379-4889 & 6.16$_{-0.80}^{+0.19}$&3.60$_{-1.60}^{+1.32}$&0.11$_{-0.03}^{+0.03}$&2.5$\pm 0.1$&176.6/121\\
    & 4889-5309& 4.38$_{-0.10}^{+0.20}$&1.39$_{-0.12}^{+0.21}$&0.12$_{-0.10}^{+0.10}$&2.4$\pm 0.1$&147.1/170\\
    & 5309-5809 & 4.12$_{-0.70}^{+0.40}$&1.60$_{-0.82}^{+0.89}$&0.4$_{-0.12}^{+0.22}$&2.1$\pm 0.2$&172.5/128\\
    & 5809-6309& 4.04$_{-0.26}^{+0.50}$&1.03$_{-0.73}^{+0.60}$&0.84$_{-0.12}^{+0.19}$&2.0$\pm 0.2$&146/166\\
    & 6309-6749& 5.29$_{-0.32}^{+0.90}$&1.35$_{-0.22}^{+0.50}$&0.98$_{-0.54}^{+0.39}$&1.7$\pm 0.1$&110/82\\
    \hline 
    \hline
    
    \end{tabular} 
            \begin{tablenotes}  
    \footnotesize              
    \item[ ]The averaged absorbed flux is derived in the 0.5--4.0 keV band using the spectra model tbabs$\times$(APEC+APEC+APEC) for the spectra of WXT(cmos36) and WXT(cmos37)+FXT. All uncertainties
quoted here correspond to a $1\sigma$ confidence level.
    \end{tablenotes}            

\end{threeparttable}  
\end{table*}

\subsection{X-ray Light Curve}\label{sec:X-ray Light Curve}

\begin{figure}[ht!]
\centering
\includegraphics[width=3.4in]{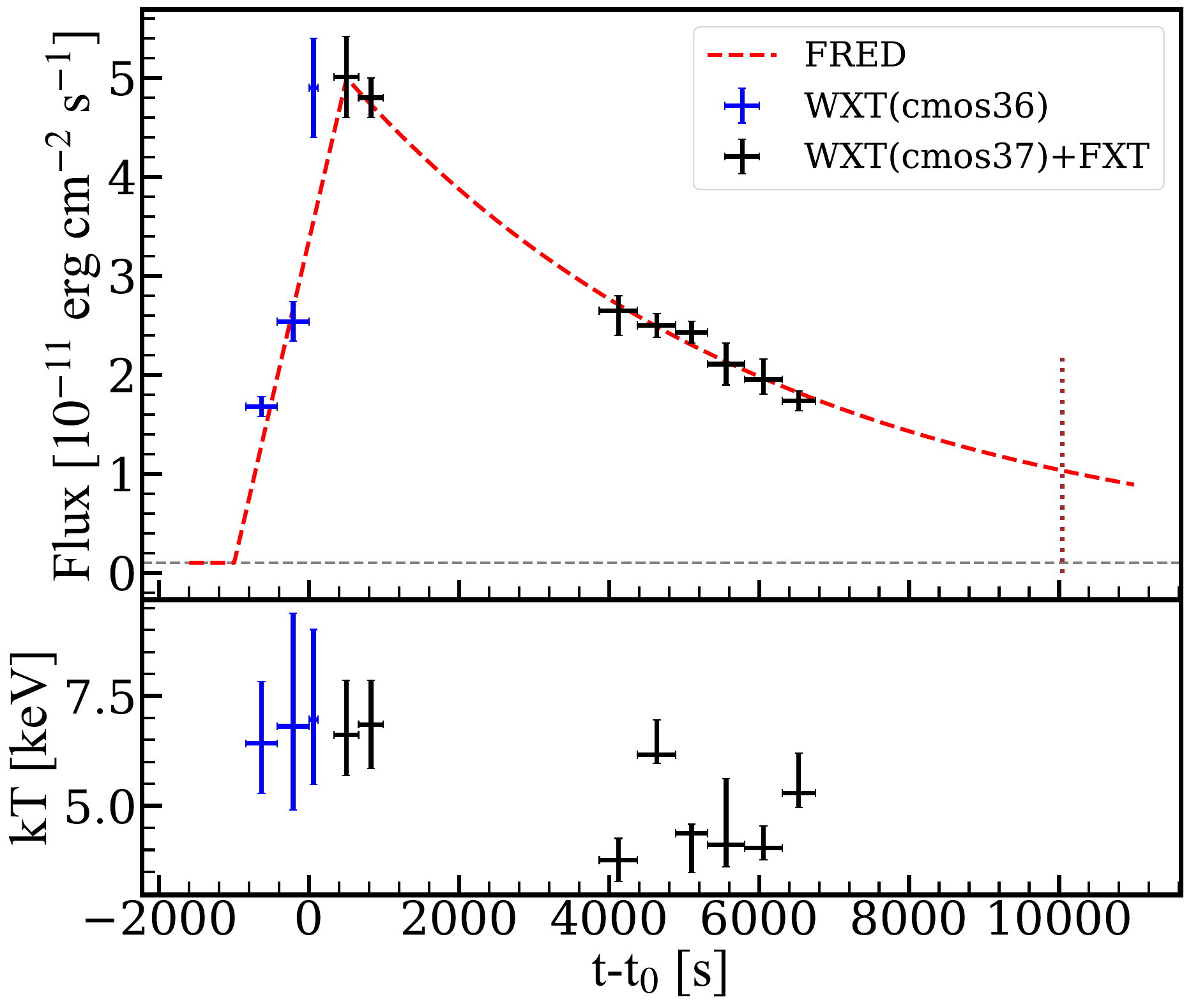}
\caption{{\it Top:} temporal modeling of the 0.5--4.0 keV X-ray flare using a FRED model. For the soft X-rays, the unabsorbed flux in the 0.5--4.0 keV band for each spectrum was estimated using the cflux model during spectral fitting in Xspec (see Section~\ref{sec:X-ray_spectral_fitting} for  the spectral fitting process.). The best-fit FRED model (red dashed line) yields a rise time ($\tau_{\text{rise}} \approx 1.4$ ks) and decay time ($\tau_{\text{decay}} \approx 5.7$ ks). The brown dotted line corresponds to the timing of the first Kast spectral observation. The gray dashed line indicates the quiescent flux during nonflaring periods. {\it Bottom:} the temperature time variation of the hot plasma component. Note: Figure~\ref{fig:Fred_curve} uses larger bins for unabsorbed flux, while Figure~\ref{fig:curve} shows 60 s binned count rates.}
\label{fig:Fred_curve}
\end{figure}

Finally, based on the time-resolved spectral analysis, we plot the X-ray light curve in Figure~\ref{fig:Fred_curve}, with the temperature variation of the hot plasma component shown in the bottom panel. The curve is consistent with a fast-rise-exponential-decay (FRED) profile characterized by a rapid rise and an exponential decay.

To quantify, we adopt the following FRED function to model the X-ray light curve~\citep{2016ApJ...832..174O, 2020yCat..51590060G,2021ApJ...910...25S,2025ApJ...980..268M}: 
\begin{equation}
    F_{\rm X}(t) = F_{\rm X,q}+ 
    \begin{cases} 
        0\, , &t<t\rm_{ST},\\
        (F_{\rm X,peak}-F_{\rm X,q})\times \frac{t-t\rm_{ST}}{t_{\rm peak}-t\rm_{ST}}, & t\rm_{ST}<t < t_{\rm peak}\, , \\
        (F_{\rm X,peak}-F_{\rm X,q})\times \exp\left(-\frac{t - t_{\rm peak}}{\tau_{\mathrm{decay}}}\right), & t > t_{\rm peak}\, ,
    \end{cases}
\end{equation}
where $t$ denotes the time since the trigger, $t\rm_{ST}$ is the time when the flux starts to increase, $t_{\rm peak}$ is the time when the flux reaches the peak, $\tau_{\mathrm{decay}}$ is the $e$-folding time of the decay, $F_{\rm X,peak}$ is the peak flux, and $F_{\rm X,q}$ is the non-flaring quiescent level flux. Since the EP observational coverage is not sufficiently long (Figure \ref{fig:curve}), the count rate and flux lightcurves do not show clearly a quiescent level. Therefore we fix $F_{\rm X,q}$ to be the historical ROSAT observed flux of $1.04\times 10^{-12}~\rm erg~s^{-1} cm^{-2}$ (Section \ref{Sec:Einstein Probe Discovery} and Table \ref{table:sources}). The best-fit model curve is shown as the dashed line in Figure ~\ref{fig:Fred_curve}. The $e$-folding decay timescale is constrained to be $\tau_{\mathrm{decay}} \approx 5.7$ ks. The rise time $\tau_{\mathrm{rise}}=t_\mathrm{peak}-t_\mathrm{ST}\approx 1.4$ ks is derived from the time difference between the flux peak and the start time.

\subsection{Flare Luminosity and Energy}

We estimate the total radiated energy $E_{\rm X}$ by integrating the flux over the duration of the flares $E_{\rm X} = 4\pi d^2\int F_{\rm X}(t)\, dt$, where $d$ is the distance to source which we adopt the reported distance of 
46.2 pc for PM J23221-031. We then obtained $E_{\rm X}=9.1 \times 10^{34}$ erg. The 0.5--4.0 keV peak flux is $F_{\rm X,peak} = 5\times 10^{-11}$ erg s$^{-1}$ cm$^{-2}$, corresponding to a peak X-ray luminosity of $L_{\rm X,peak} = 1.3 \times 10^{31}$ erg s$^{-1}$.

\begin{figure}[ht!]
    \centering
    \subfigure{
        \includegraphics[width=3.6in]{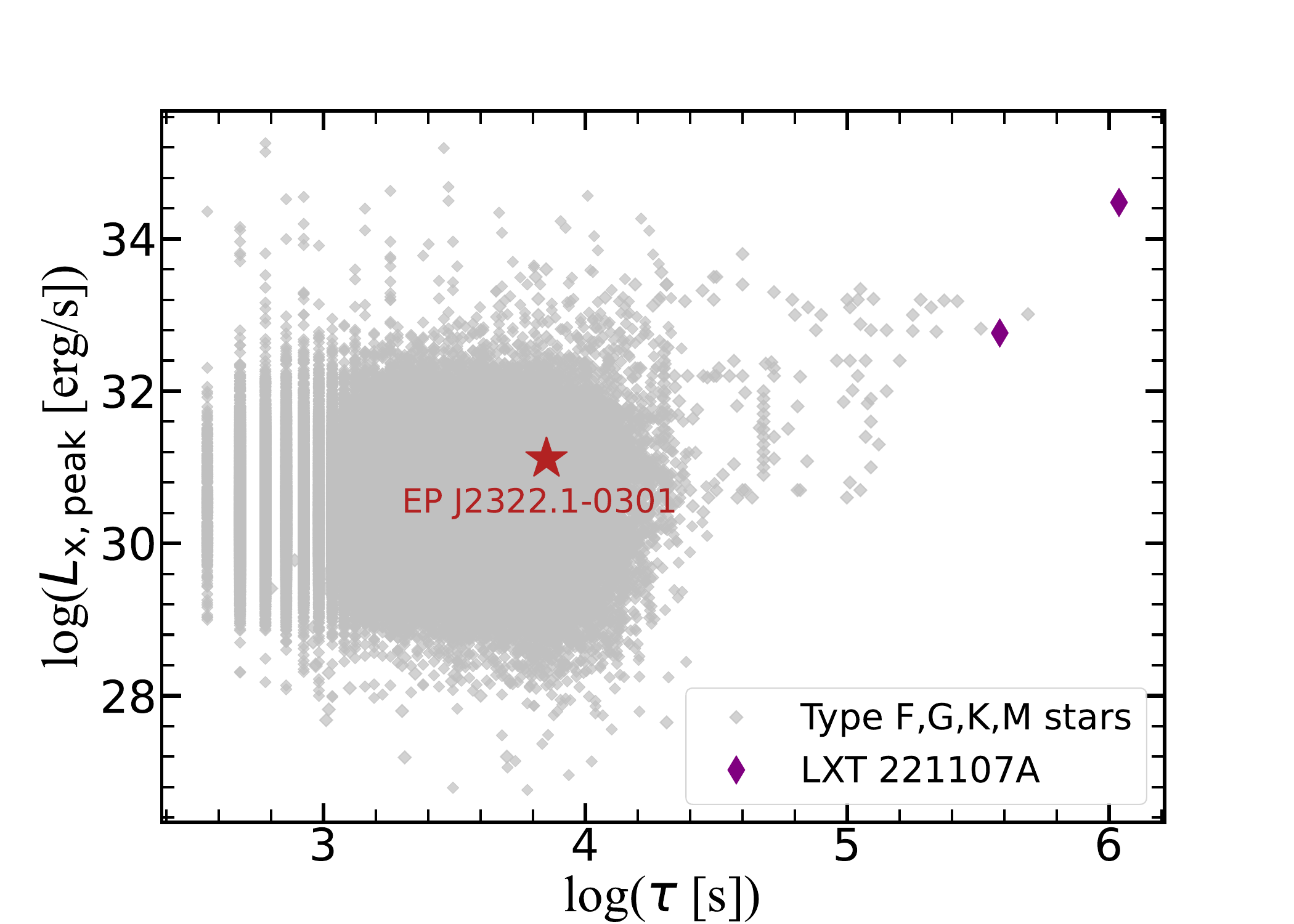}
    \label{fig:t-L}}  
    \caption{Flare duration versus flare peak luminosity for EP J2322.1-0301 and a sample of stellar X-ray flares. The gray squares are for flares from stars of different spectral types~\citep{2023A&A...669A..15Y,2021ApJ...920..154G,2015A&A...581A..28P,1989PASJ...41..679T,1997A&A...328..565E,2001A&A...375..196F,2012MNRAS.419.1219P,2016PASJ...68...90T,2021ApJ...910...25S,2023MNRAS.518..900K}. The two purple diamonds denote the most energetic and longest-duration stellar flare on record, LXT 221107A, corresponding to the main flare and secondary flare, respectively ~\citep{2025ApJ...980..268M}.}
    \label{fig:T_E_L}
\end{figure}

In Figure~\ref{fig:T_E_L} we plot the flare peak luminosity versus the flare duration $\tau$, defined as the sum of the rise time $\tau_\mathrm{rise}$ and decay duration $\tau_\mathrm{decay}$, for EP J2322.1-0301 along with a sample of stellar X-ray flares ~\citep{2023A&A...669A..15Y,2021ApJ...920..154G,2015A&A...581A..28P,1989PASJ...41..679T,1997A&A...328..565E,2001A&A...375..196F,2012MNRAS.419.1219P,2016PASJ...68...90T,2021ApJ...910...25S,2023MNRAS.518..900K}. Notably, the peak luminosity of EP J2322.1-0301 falls within the main region of
the parameter space of previously documented stellar X-ray flares.

\subsection{H$\alpha$ Emission}
Here we analyze the H$\alpha$ emission in the \textbf{$t=2.79$ hr} spectrum (Figure~\ref{fig:KAIT_spectrum}). The equivalent width (EW) of this spectral feature, EW = $\int_{\lambda} ({F_\lambda-F_{\rm cont}})/{F_{\rm cont}}\, d\lambda$, where $F_\lambda$ denotes the wavelength-dependent flux of the emission line and $F_{\rm cont}$ represents the corresponding continuum flux density, is estimated to be 5.07 $\text{\AA}$. Then the H$\alpha$ line flux is derived as $F_{{\rm H}\alpha}$ = EW $\times F_{\rm cont} = 7.6 \times 10^{-13}$ erg s$^{-1}$ cm$^{-2}$. 

Note that~\cite{2024ApJ...961...23N} derived an empirical power-law scaling relation between the energies emitted in X-rays (0.1-100 keV) and $\rm H\alpha$ lines for stellar flares. Here we estimate the energy ratio $E _\mathrm{X}/E_\mathrm{{H\alpha}}$ for EP J2322.1-0301. The $\rm H\alpha$ spectra of EP J2322.1-0301 were obtained after the peak of the soft X-ray emission, when the flare had entered its decay phase. It has been noted in the literature~\citep{2002SoPh..208..297V,1988A&A...193..229D, 2011A&A...534A.133F,2020PASJ...72...68N,2022PASJ...74..477K} that for solar \& stellar flares both the soft X-ray and $\rm H\alpha$ emissions follow an exponential decay law, and the temporal profiles in the two wavelengths have about the same peak times and the same decay timescales. 

Therefore, we estimate the energy ratio to be the flux ratio at the $\rm H\alpha$ observation time: $E _\mathrm{X}/E_\mathrm{{H\alpha}}=F\rm_ X(t=2.79~hr)/F\rm_{H\alpha}(t=2.79~hr)=15.2$. Here to get $F\rm_ X(t=2.79~hr)$ we have extrapolated from $F\rm _{X,peak}$ using the best-fit FRED profile (Figure~\ref{fig:Fred_curve}). A conversion factor of 1.2  is also included for extrapolating $F\rm_X$ from the 0.5-4.0 keV band to the 0.1-100 keV. This value falls well within the range reported in previous studies, with ~\cite{2022PASJ...74..477K} reporting an $E_\mathrm{X}/E_{\rm H\alpha}\approx 10$, and ~\cite{2024ApJ...961...23N} finding a broader range of 10 to 100.

\subsection{Stellar Rotation}
\label{sec:Stellar Rotation}

The star PM J3221-0301 rotates with a period of 1.28 days~\citep{2013AcA....63...53K}. If a flare happens in a region close to the stellar limb, the flaring region may gradually move into the hemisphere facing away from the observer~\citep{1999A&A...344..154S,2003ApJ...582..423G,2024LRSP...21....1K}. Such rotational modulation can lead to a gradual decline in the observed emission, and thus may give an apparently shorter decay timescale relative to the intrinsic evolution~\citep{2012MNRAS.419...29J}.

Due to the sparsity of the data coverage, we can not rule out this possibility at working for EP J2322.1-030. Given the rotation period of 1.28 days, at $t_0+30.53$ hr the flaring region should have completed nearly a full rotation and reappeared in the observer's line of sight. If the flare were still active at that time, significant $\rm H\alpha$ emission should still be detectable. However, no enhanced $\rm H\alpha$ feature is observed (Figure~\ref{fig:KAIT_spectrum}). This indicates that the flare had fully ceased by then. Therefore, we can place an upper limit of 30.53 hr on the intrinsic flare duration, if the observed decay in EP J2322.1-030 were due to the rotational modulation.

\section{Conclusion}\label{sec:conclusion and discussion}

In this work, we present the discovery and multiwavelength characterization of the X-ray transient EP J2322.1-0301, discovered by the Einstein Probe (EP) on 2024 September 27. We identify it as a stellar flare from the high proper motion K-type star PM J23221-0301. The association with a stellar flare is confirmed by multiple lines of evidence: the spatial coincidence between the X-ray transient and PM J23221-0301 (within $10''$ position uncertainly), the X-ray light curve shape consistent with a FRED profile, the transient H$\alpha$ emission line in optical spectra, and a consistency of flare energetics with known stellar-flare properties.

The FRED-profile fit to the X-ray light curve of EP J2322.1-0301 suggests a rise time $\tau_\mathrm{rise}\approx 1.4$ ks and a decay time $\tau_\mathrm{decay}\approx 5.7$ ks, yielding a total duration of $\sim 7.1$ ks. These timescales fall within the typical range of stellar flares reported in the literature  \citep{2015A&A...581A..28P,1989PASJ...41..679T,1997A&A...328..565E,2001A&A...375..196F,2012MNRAS.419.1219P,2016PASJ...68...90T,2016AJ....152..168L,2023MNRAS.518..900K,2021MNRAS.505L..79Y}.

Optical spectra obtained with the Kast spectrograph at Lick Observatory provide critical evidence for the flare association: a prominent H$\alpha$ emission line detected in the spectrum taken 2.79 hr post-trigger disappeared in the follow-up observation 1.27 days later. The 2-hr time interval between the X-ray peak and the $\rm H\alpha$ detection is significantly shorter than the typical recurrence interval of 10 hr-30 days for stellar flares~\citep{1999sf99.proc..312I,2003SoPh..215..127K}. We therefore consider the probability that the observed $\rm H\alpha$ enhancement is caused by another independent flare to be low, though it cannot be entirely ruled out.

We compare EP J2322.1-0301 with stellar flares observed in known stars, as shown in Figure ~\ref{fig:T_E_L}. Its estimated total energy release at a distance of 46.2 pc (${E_{\rm X} = 9.1 \times 10^{34}}$ erg) falls within the typical range observed for stellar flares. Such a comparison provides valuable context for understanding the scaling relations between stellar-flare energy and duration.

The spectral analysis reveals a multitemperature plasma, which may suggest a possible stratification of the flaring plasma. This finding is consistent with the standard flare-loop models, where chromospheric evaporation generates hotter plasma in the upper regions of the loop and cooler material at lower altitudes~\citep{1997A&A...325..782R,1998A&A...334.1028R,1985ApJ...289..414F}. This inclusion of an additional component resolves the low-energy residuals observed in the two-temperature fit (Figure~\ref{fig:spec_res_fit}); this supports that a multitemperature stratified plasma exists in the flare process, reflecting the rapid heating and gradual cooling processes triggered by magnetic reconnection.

Assuming that the temporal evolution of the $\rm H\alpha$ line emission tracks that of the X-rays, we derived the total X-ray-to-$\rm H\alpha$ energy ratio of $E_{\rm X}/E_{{\rm H}\alpha}\approx15.2$, consistent with the empirical X-ray vs. $\rm H\alpha$ relations observed in stellar flares~\citep{2022PASJ...74..477K, 2024ApJ...961...23N}. We can not rule out a possibility that the observed X-ray decay phase is caused by the star's rotational modulation to a longer flare that happens close to the stellar limb. If that were the case, we can put an upper limit of 30.53 hr on the intrinsic flare duration (Section~\ref{sec:Stellar Rotation}). 

This study is one of the early achievements in the synergistic detection of K-type star flares using WXT and FXT of EP. EP identified the transient EP J2322.1-0301 via wide-field monitoring in the 0.5-4.0 keV energy band using WXT, achieving a localization error of only 2.461 arcminutes. Subsequently, FXT performed rapid focusing within 329 s, reducing the localization error to $\sim 10''$, which is crucial for identifying optical counterparts. Ultimately, a precise spatial correlation was established between the X-ray source and the optical star PM J23221-0301, with a position offset of less than 6 arcseconds. This ``first detection, then focusing'' observational paradigm offers unique advantages for the rapid identification of short-timescale transient flares (e.g., stellar flares) and the investigating their physical associations. It provides a methodological reference for future large-sample statistical studies of stellar flares based on EP. EP's ability to detect and precisely localize such transient events highlights its capability for studying stellar flares. 

\section*{Acknowledgements}
We thank the referee for valuable comments which improves our manuscript. This work is supported by the National Natural Science Foundation of China (grant NSFC-12393814). C.G. acknowledges support from 
NSFC grants 12373007 and 12422302.
A.V.F.’s research group at UC Berkeley acknowledges financial assistance from the Christopher R. Redlich Fund, Gary and Cynthia Bengier, Clark and Sharon Winslow, Alan Eustace (W.Z. is a Bengier-Winslow-Eustace Specialist in Astronomy), William Draper, Timothy and Melissa Draper, Briggs and Kathleen Wood, Sanford Robertson (T.G.B. is Draper-Wood-Robertson Specialist in Astronomy), and numerous 
other donors.

KAIT and its ongoing         
operation were made possible by donations from Sun Microsystems, Inc.,        
the Hewlett-Packard Company, AutoScope Corporation, Lick Observatory,        
the U.S. NSF, the University of California, the Sylvia\&Jim                  
Katzman Foundation, and the TABASGO Foundation.
A major upgrade of the Kast spectrograph on the Shane 3 m telescope at Lick Observatory, led by Brad Holden, was made possible through gifts 
from the Heising-Simons Foundation, William and Marina Kast, and the 
University of California Observatories.
We thank the staff at 
Lick Observatory for their assistance.  Research at Lick Observatory          
is partially supported by a generous gift from Google.  

\bibliographystyle{elsarticle-harv} 
\bibliography{reference}
\end{CJK*}
\end{document}